\documentstyle[twoside,fleqn,espcrc2,epsfig]{article}

\newcommand{\bbox}[1]{{ \mbox{\boldmath $#1$}}}
\newcommand{\tr}{\mathop{\mathrm{tr}}}
\renewcommand{\d}{\mathop{\mathrm{d}}}
\newcommand{\xis}{{\xi_{\mathrm{s}}}}

\newcommand{\chis}{{\chi_{\mathrm{s}}}}
\newcommand{\etas}{{\eta_{\mathrm{s}}}}
\newcommand{\betas}{{\beta_{\mathrm{s}}}}
\newcommand{\betac}{{\beta^{\mathrm{c}}}}

\newcommand{\kappas}{{\kappa_{\mathrm{s}}}}
\newcommand{\Orth}[1]{\mathrm{O}(#1)}

\title{Critical exponents and unusual properties of the broken phase\\
       in the 3d-RP$^2$ antiferromagnetic model.\thanks{
        Presented by V.~Mart\'{\i}n-Mayor.
Partially supported by CICyT \hbox{AEN93-0604-C03} and AEN95-1284-E, Spain.
        \hbox{e-mail}: \{hector,laf,victor,sudupe\}@lattice.fis.ucm.es .}}

\author{H.G.~Ballesteros, L.A.~Fern\'andez, V.~Mart\'{\i}n-Mayor
        and A.~Mu\~noz~Sudupe\\
        \medskip
        Depto. de F\'{\i}sica Te\'orica I,
        Fac. de CC. F\'{\i}sicas,
        Univ. Complutense de Madrid, 28040 Madrid, Spain.}

\begin{document}

\begin{abstract}
We present the results of a Monte Carlo simulation of the
antiferromagnetic RP$^2$ model in three
dimensions. With finite-size scaling techniques we accurately measure
the critical exponents and compare them with those of O($N$)
models. We are able to parameterize the
corrections-to-scaling.  The symmetry properties of the broken phase
are also studied.
\end{abstract}

\maketitle

\section{Introduction.}

AntiFerromagnetic (AF) models exhibit interesting characteristic
properties. For instance, AF couplings on non-bipartite
lattices, as the triangular Ising model, produce frustration. On
bipartite lattices they can produce disordered Ground States, as for the
AF three states Potts Model in three dimensions, which
belongs to the XY Universality Class. It is then possible to find new
Spontaneous Symmetry Breaking (SSB) patterns, and maybe new
Universality Classes. Our interest in these models is twofold:
they apply to Condensed Matter systems (spin glasses, helical and
canted spin systems,${^3}\mathrm{He}$ superfluid phase
transition, etc.) and four dimensional Field Theory, as they might
offer some insight on the formulation of non asymptotically-free
interacting theories.
For the AF RP$^2$ model, the finite-size scaling analysis close to the
transition, strongly suggests that the action's
$\mathrm{O}(3)$-symmetry is broken, yielding an SO(3)/\{1\} SSB
pattern. The perturbative prediction for such an SSB pattern is that
the transition can either be on the O(4) Universality Class or that it
should be first order or tricritical~\cite{AZARIA}.

\section{The model.}

We consider a system of three components, normalized vectors, placed
on the nodes of a cubic lattice. They interact through a first neighbors
coupling

\begin{equation}
S=\beta\sum_{<i,j>}{(\bbox{\sigma}_i\cdot\bbox{\sigma}_j)}^2\ .
\end{equation}

This action presents a global $\Orth{3}$ symmetry, and a local Z$_2$
one ($\bbox{\sigma}_i\rightarrow -\bbox{\sigma}_i$). From
Elitzur's theorem, it follows that we are really studying RP$^2$
variables.  For positive $\beta$, this model suffers a first order
phase transition, all spins aligned or anti-aligned with an arbitrary
direction. This corresponds to the nematic phase of liquid
crystals. At $\beta\approx -2.41$ the system undergoes a second order
phase transition~\cite{SHROCK,RP2}. The symmetry description is however,
more complicated. Let us call the spin placed on $(x,y,z)$ even or
odd, according with the parity of $x+y+z$. At $\beta=-\infty$, the
ground state consists on all the spins on the, for example, even
sublattice, aligned or anti-aligned with an arbitrary
direction, while odd spins lie randomly on the perpendicular plane.
This state has a remaining $\mathrm{O}(2)$-symmetry, but the symmetry
between even-odd sublattices is broken. However, thermal fluctuations
induce an interaction between the spins on the plane sublattice, as an
alignment allows stronger fluctuations on the other sublattice.

As the natural variable is the tensorial product of $\bbox{\sigma}$ by
itself, we consider the traceless tensorial field 
\begin{equation}
\mathrm T_i^{\alpha\beta}=
 \sigma_i^\alpha \sigma_i^\beta - \frac{1}{3}\delta^{\alpha\beta}\ .
\end{equation}

We define the normalized non-staggered and staggered magnetizations
as
\begin{eqnarray}
{\bf M}_{\hphantom{\mathrm{s}}}&=&\frac{1}{V}\sum_{x,y,z} 
{\mathbf T}_{(x,y,z)} ,\\
{\bf M}_{\mathrm{s}}&=&\frac{1}{V}\sum_{x,y,z} (-1)^{x+y+z} 
{\mathbf T}_{(x,y,z)}.
\end{eqnarray}

In the simulation we actually measure
\begin{equation}
M=\left\langle \sqrt{\tr {\mathbf M}^2}\right\rangle\ ,
\label{MAG}
\end{equation}
and the associated susceptibility
\begin{equation}
\chi=V \left\langle \tr {\mathbf M}^2\right\rangle\ .
\end{equation}

We have found very useful the measure of the second momentum
correlation length defined as
\begin{eqnarray}
\xi=\left(\frac{\chi/F-1}{4\sin^2(\pi/L)}\right)^{1/2},
\end{eqnarray}
where $F$ is the Fourier transform of the correlation function at minimal
non zero momentum, as well as the scaling function (related to Binder's 
parameter)
\begin{eqnarray}
\kappa=\frac{\langle \left(\tr {\mathbf M}^2\right)^2\rangle}
              {\langle \tr {\mathbf M}^2\rangle^2} .
\label{BINDER}
\end{eqnarray}

The staggered counterparts of the quantities 
\hbox{(\ref{MAG}-\ref{BINDER})}
are defined analogously.

We have also measured the first and second neighbors energies:
\begin{eqnarray}
E_1&=& \frac {1}{3V} \sum_{<i,j>} (\bbox{\sigma}_i\cdot\bbox{\sigma}_j)^2 ,\\
E_2&=& \frac {1}{6V} \sum_{\ll i,j\gg} (\bbox{\sigma}_i\cdot\bbox{\sigma}_j)^2.
\end{eqnarray}

$E_1$ has been used for the Ferrenberg-Swendsen 
extrapolation method, and for calculating $\beta$-derivatives. $E_2$
is useful to obtain information about the structure of the broken 
phase.

\section{The simulation.}

We have used a Metropolis update. Cluster methods
are feasible but not efficient. 

We have measure the
integrated and exponential autocorrelation times for $E_1$, $\chi$ and
$\chis$. The integrated times satisfy $\tau_{E_1}^{\mathrm{int}}<
\tau_{\chi}^{\mathrm{int}}<\tau_\chis^{\mathrm{int}}$
(for instance, in the larger lattice the ratio is
1:3:7). However, the exponential times are almost equal, and they 
are near $\tau_\chis^{\mathrm{int}}$. So, we have confidence that
the latter is the larger autocorrelation time. Further details
can be found in refs. \cite{RP2}.

In table \ref{MC} we present the number of Monte Carlo iterations
performed at each lattice size, together with the corresponding
autocorrelation time.  The expected behavior $\tau\propto L^2$ is well
satisfied.

\begin{table}[h]
\caption{Number of Monte Carlo sweeps performed for different lattice
 sizes.  Measures have been taken every 10 sweeps.  We have discarded
 in each case about $200\tau^{\mathrm{int}}_{\chis}$ iterations for
 thermalization.}
\begin{tabular*}{\hsize}{@{\extracolsep{\fill}}rccrc}
\hline
\hline 
$L$ & MC sweeps($\times 10^6$) &$\tau^{\mathrm{int}}_{\chi_{\mathrm {s}}}$ & \# of $\tau^{\mathrm{int}}_{\chi_{\mathrm {s}}}$\\\hline
6       & 6.71   &  7.37(3)  &910,000   \\    
8       & 17.07  & 11.41(4)  &1,496,000 \\    
12      & 6.51   & 24.9(2)   &261,000  \\    
16      & 22.14  & 44.5(3)   &498,000  \\    
24      & 8.77   &107(3)     &82,000  \\    
32      & 28.51  &175(6)     &163,000  \\    
48      & 3.93   &410(20)    &9,600 \\\hline
\hline
\end{tabular*}
\label{MC}
\end{table}

\section{Finite size scaling techniques.}

For an operator, $O$, with  critical exponent $x_O$, we have
\begin{eqnarray*}
\langle O(L,\beta) \rangle=L^{x_O/\nu}
\left(F_O(\xi(L,\beta)/L)
+O(L^{-\omega})\right).
\end{eqnarray*}
Order ${\xi(\infty,\beta)}^{-\omega}$ terms, are negligible in the
critical region. 

Measuring at two lattice sizes $L$ and
$sL$ and computing the quotient 
\begin{equation}
Q_O=\frac{\langle O(sL,\beta)\rangle}{\langle O(L,\beta)\rangle}\ ,
\end{equation}
we can eliminate the scaling function $F_O$ just by choosing the $\beta$
value such that the ratio between correlation lengths is $s$,
obtaining
\begin{equation}
\left.Q_O\right|_{Q_\xi=s}=s^{x_O/\nu}+O(L^{-\omega})\ .
\label{QUOTIENT}
\end{equation}

\section{Exponents}

To compute the $\nu$ exponent we consider operators like
the $\beta$ derivatives of the correlation length or the
logarithm of the magnetization (\hbox{$x_{\d \xi/\d \beta}=\nu+1$},
$x_{\d \log M/\d \beta}=1$). In table \ref{EXPO} we
present in the second column the results obtained from
$\left.Q_{\d \xis/\d \beta}\right|Q_{\xis}=2$. The values
using other operators involving $\xi$ are similar but slightly
worse. In the case of operators involving magnetizations, the
statistical errors are smaller, but the finite-size effects
are greater. We do not find any significant deviation between
the staggered and non-staggered channels what supports the
equality $\nu=\nu_{\mathrm{s}}$ necessary to define a single
continuum limit.

In the case of magnetic exponents, we expect a different behavior for
the staggered and non-staggered channels.  We obtain the respective
$\etas$ and $\eta$ exponents from $\gamma_{\mathrm (s)}/\nu$ or
$\beta_{\mathrm(s)}/\nu$ using the scaling relations
$\eta_{\mathrm(s)}=2-\gamma_{\mathrm(s)}/\nu$ or
\hbox{$\eta_{\mathrm(s)}=2-d+2\beta_{\mathrm(s)}/\nu$}.  The quality of the
results depends on the observable measured.  With a good selection of
both the $O$ operator and the definition of correlation length the
corrections-to-scaling terms can be largely reduced. We should remark
that the operator used to measure the correlation length in practice
can be any quantity that scales linearly with $L$ at the critical point, 
such as $\kappa L$.  
However, the more interesting property of the method we use
is that even if the operator $O$ is a rapidly varying function of the
coupling at the critical point, as the magnetization or the
susceptibility are, the statistical correlation between $Q_O$ and
$Q_\xi$ allows a very precise measure of the critical exponents.

In table \ref{EXPO}, (col. 3), we display the results for $\etas$ 
using $\chis$ as operator and $\xis$ as correlation length.
Regarding $\eta$ exponent the use of $\xi$ or $\xis$ as
correlation length produce large corrections-to-scaling. We have
realized that those effects are strongly reduced when using
$\kappa L$ as the {\it correlation length} operator (see col. 4).

\begin{table}[t]
\caption{ $\nu$, $\etas$ and $\eta$ exponents obtained from
(\ref{QUOTIENT}) using as operators $\d \xis/\d \beta$, $\chis$ and
$\chi$ respectively. As correlation length we take $\xis$ in all cases
but the latter where we have used $\kappa L$. The pairs are of type
$(L,2L)$.  }
\begin{tabular*}{\hsize}{@{\extracolsep{\fill}}rlll}
\hline
\hline
$L$  &\multicolumn{1}{c}{$\nu$}
     &\multicolumn{1}{c}{$\etas$}
     &\multicolumn{1}{c}{$\eta$}\\
\hline
6  & 0.786(6) & 0.0431(10)& 1.332(6) \\
8  & 0.785(4) & 0.0375(7) & 1.332(4) \\
12 & 0.789(8) & 0.0357(17)& 1.321(13)\\
16 & 0.786(6) & 0.0375(12)& 1.340(5) \\
24 & 0.77(2)  & 0.038(5)  & 1.334(18)\\
\hline
\hline
\end{tabular*}
\label{EXPO}
\end{table}

\section{Scaling corrections}

At very large volume, scaling functions for different lattices, such as
$\xis/L$, $\xi/L$, $\kappa$ or 
$\kappas$ should  cross at $\beta^{\mathrm{c}}$.
Scaling corrections shift the crossing point, corresponding to
a pair of lattices \hbox{$(L_1,L_2=sL_1)$}, an amount given by
\begin{equation}
\Delta\beta^{L_1,L_2}\propto
\frac{1-s^{-\omega}}{s^{\frac{1}{\nu}}-1}L_1^{-\omega-\frac{1}{\nu}}\ .
\label{SHIFT} 
\end{equation}

As the crossing point of correlation lengths and Binder-like parameters 
tend to the critical point from opposite sides, it is sensible to
perform a global fit, with the full covariance matrix to take into
account the statistical correlation of all the data. 
We have fitted our data either fixing the small lattice, $L_1$, or an $s$
value. For instance, for $L_1=8$, we obtain
\begin{equation}
\begin{array}{lcr}
\betac&=&-2.4085(3)\ ,\\
\omega&=&0.86(4)\ ,\\
\chi^2/\mathrm{d.o.f.}&=&12.2/14\ .
\end{array}
\end{equation}
Using also the $L_1=6$ data or fixing $s=2$ the results are compatible
(see ref.~\cite{RP2}).

After measuring the, universal, corrections-to-scaling exponent $\omega$
we can estimate the finite-size effects on the critical exponents
using 
\begin{equation}
\frac{x}{\nu}-
\left.{\frac{x}{\nu}}\right|_{(2L,L)}\propto L^{-\omega}\ .
\end{equation}

From the results quoted in table \ref{EXPO} we conclude that the
values for $\nu$ are rather stable when increasing the lattice size,
but a proper accounting of finite-size effects is crucial in the case
of $\eta$ exponents for some operators (see also ref.~\cite{ON} where
this method is applied to three dimensional O($N$) models).

Anyway, the presence of an $L$ dependence, even when it is not
measurable, requires an increasing of the error bars for a safe
determination of the systematic error due to finite-size effects.
We summarize the results giving the value for the (16,32) pair, 
with a second error bar
that corresponds to the increasing of the error due to the
infinite volume extrapolation,
\begin{equation}
\begin{array}{lcl}
\nu   &=&0.783(5)(6)\ ,\\ 
\eta_{\mathrm{s}} &=&0.0380(12)(14)\ ,\\
\eta  &=&1.339(5)(5)\ .
\end{array}
\label{EXPOINFTY}
\end{equation}
Some minor differences between the values in table \ref{EXPO} and 
(\ref{EXPOINFTY}) exist because here we average between the results
of several operators not displayed in the table.

The exponent $\nu$ is two standard deviations apart from the value for
the three-dimensional O(4) model ($\nu_{\Orth4}=0.755(8)$\cite{ON}) what
supports that the AF transition in the RP$^2$ model belongs to a new
Universality Class, but the opposite cannot be completely discarded.

\section{Vacuum structure}

In the $\beta=-\infty$ limit, there is a breakdown of the symmetry
between the odd and even sublattices, but a global $\Orth2$ symmetry
still remains. At the critical region, the presence of fluctuations
can change this picture. 

To study the structure of the broken phase near the critical point
we have analyzed the FSS of two operators
\begin{eqnarray}
A&=&\langle (\tr {\mathbf M}_{\mathrm{s}} {\mathbf M})^2\rangle,\\
B&=&\langle \tr [{\mathbf M}_{\mathrm{s}}, {\mathbf M}][
                [{\mathbf M}_{\mathrm{s}}, {\mathbf M}]^\dagger\rangle.
\end{eqnarray}

The $A$ operator depends on the difference of the magnetizations
squared of both sublattices ($\tr {\mathbf M}_{\mathrm even}^2-\tr
{\mathbf M}_{\mathrm odd}^2$), so, if the even-odd symmetry were not
broken it should scale as $L^{-2d}$. The obtained value
\begin{equation}
x_A/\nu=-3.389(15)\ ,
\end{equation}
agrees with $-2(\beta+\betas)/\nu$ and discards a remaining even-odd
symmetry.

The $B$ operator measures if the tensors $\mathbf M$ and ${\mathbf
M}_{\mathrm s}$ commute (what is equivalent to the commutation of
${\mathbf M}_{\mathrm even}$ with ${\mathbf M}_{\mathrm odd}$). If
they did not, a simultaneous diagonalization is not possible, and a
remaining $\Orth2$ symmetry is discarded. The expected FSS for an
unbroken $\Orth2$ symmetry correspond to
$x_B/\nu=-d-2\betas/\nu=4.04(2)$ while the value obtained numerically
is
\begin{equation}
x_B/\nu=-3.406(11)\ .
\end{equation}

To check if both breakdowns occur deep in the broken phase, we have
studied the second neighbor energy $E_2$ for each sublattice, as well
as the eigenvalues of the magnetization tensors.

The analysis of the $E_2$ histograms show an unambiguous signature of
breakdown of the even-odd symmetry along all the broken phase.  It is
interesting to remark that the energy of the less aligned sublattice
takes for $\beta\to-\infty$ the value 0.532(5) to be compared with the
$\beta=-\infty$ value, 0.5.  The presence of fluctuations acts as an
effective coupling that favors an alignment.

The intensity of that induced interaction increases when approaching
to the critical point. To know where it actually produces a breakdown
of the $\Orth2$ symmetry, we have studied the eigenvalues of the
$\mathbf M$ matrix. We have observed that the difference between the
two smaller eigenvalues goes to zero when the lattice size grows, for
$\beta\le -4$. At $\beta=-3$ a nonzero value is not completely
excluded but we would need lattice sizes larger than 48 to obtain a
reliable measure.  In order to clarify this point, we project to study
an extended model where an explicit second neighbor coupling is added.

\end{document}